\documentstyle[epsfig,epsf]{article}

\begin{document}
\large
\begin{flushright}
DFUB 2001-09  \par
Bologna, \today
\end{flushright}

\begin{center}
{\Large {\bf ATMOSPHERIC NEUTRINO OSCILLATIONS IN MACRO}}
\end{center}

\vskip .7 cm

\begin{center}

GIORGIO GIACOMELLI and MIRIAM GIORGINI \par
for the MACRO Collaboration{\footnote {For the full list of 
the Collaboration see the last paper of ref. \cite{macro}}}

\par~\par 

{\it Dept. of Physics, University of Bologna,
  V.le C. Berti Pichat 6/2\\
 Bologna, I-40127, Italy\\} 

E-mail: giorgio.giacomelli@bo.infn.it , miriam.giorgini@bo.infn.it 

\par~\par

 Invited paper at NO-VE, Int. Workshop on Neutrino 
Oscillations in Venice, \\ Venice, Italy, 24-26 July 2001. 

\vskip .7 cm
{\bf \normalsize Abstract}\par
\end{center}

{\normalsize In this paper we first give a short overview of the MACRO detector
which was operational at the Gran Sasso Laboratory from 1989 till the
end of 2000. Then 
we present and discuss the results on atmospheric muon neutrino oscillations,
 concerning medium ($\sim 4$ GeV) and high ($\sim 50$ GeV) energy data. Using
 the
Multiple Coulomb Scattering of muons inside the lower part of the detector,
 estimates of neutrino energies were made for the high energy sample. The
data on angular distributions, absolute flux and $L/E_\nu$
distributions favour $\nu_\mu \longleftrightarrow \nu_\tau$ oscillations with
maximal mixing and $\Delta m^2=2.5 \cdot 10^{-3}$ eV$^2$.} 
   
\vspace{5mm}

\section{Introduction}
MACRO was a large multipurpose underground detector designed primarely
to search for rare particles in the penetrating cosmic radiation. Though
it was optimized to search for the supermassive magnetic monopoles predicted
by Grand Unified Theories (GUT) of the electroweak and strong 
interactions, the MACRO research program included a wide range of topics 
in areas of astrophysics, nuclear and particle physics and cosmic ray 
physics. In particular it planned to study atmospheric neutrinos 
via the detection of
upgoing muons produced in charge current (CC) interactions of $\nu_\mu$
in the rock below the detector and inside its lower part. 

\begin{figure}
\begin{center}
\mbox{\epsfig{figure=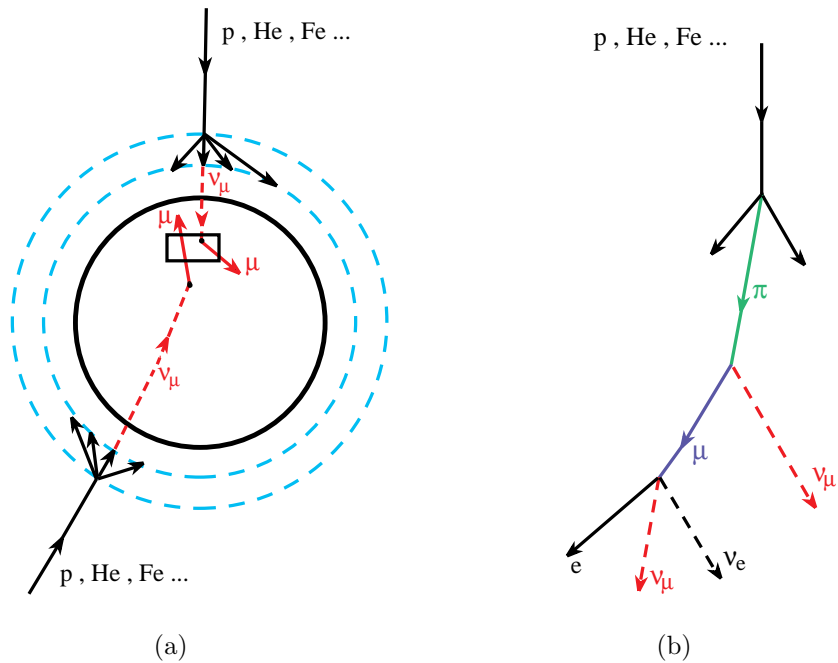,width=11cm}}
\begin{center}
{\normalsize (a) \hspace{6cm} (b)} 
\end{center}
\vspace{-2mm}
\caption{ (a) Illustration of the production, travel and interactions
of atmospheric muon neutrinos; (b) interaction of a primary
cosmic ray, production of pions (and kaons) and their decays 
leading to the atmospheric $\nu_e,~\nu_\mu$.}
\label{fig:catena}
\end{center}
\end{figure}

A high energy primary cosmic ray, proton or nucleus, interacts 
in the upper atmosphere producing a large number of pions 
and kaons, which decay yielding  muons and muon
neutrinos; also the muons decay yielding muon and electron 
neutrinos, Fig:~\ref{fig:catena}.
 The ratio of the
numbers of muon to electron neutrinos is about 2, 
 and $N_{\nu}/N_{\overline\nu} \simeq 1$.
 These neutrinos are produced in a spherical
surface at about 20 km above ground and  they proceed at 
high speed towards the Earth. An underground detector is ``illuminated" by
a flux of neutrinos from all directions.

 At low energies, $E_\nu \sim 1$ GeV, the numbers of atmospheric 
neutrinos predicted by different authors differ by about 
$20\div 30\%$~\cite{nulow};
 at higher energies, $E_\nu > 10$ GeV, the predictions are more
reliable, with an estimated systematic uncertainty of about 
$15\%$~\cite{nuhigh}, 
 almost one half of the uncertainty at low energies. However the predicted
relative rates of $\nu_\mu$ to $\nu_e$ and the shapes of the zenith angle 
distributions are affected by 
considerably lower systematic errors. Other sources of systematic
uncertainties arise from the knowledge of the 
neutrino-nucleon cross sections and from the propagation of muons and
electrons in different materials.

Several large underground detectors studied atmospheric neutrinos, mainly 
via charged current 
interactions, where a $\nu_\mu$ gives rise to a $\mu^-$, and thus to a track,
 and a $\nu_e$ yields an $e^-$ and thus to an electromagnetic shower. 

\begin{figure}
\begin{center}
\mbox{\epsfig{figure=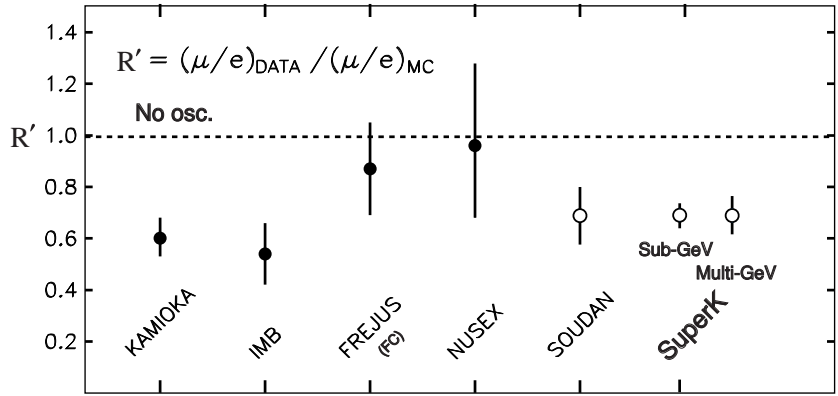,width=11cm}}
\caption{Double ratios $R^\prime$
measured by several atmospheric neutrino experiments.}
\label{fig:ratios}
\end{center}
\end{figure}

The early water Cherenkov detectors IMB \cite{imb}
and Kamiokande \cite{kamioka} reported 
anomalies in the double ratio of muon to electron
neutrinos integrated over all zenith angles, while the tracking 
calorimeters NUSEX \cite{nusex} and 
 Frejus \cite{frejus}, and the Baksan \cite{baksan} scintillator detector
did not find any deviation, see Fig:~\ref{fig:ratios}.

Later the Soudan 2 \cite{soudan}, MACRO \cite{macro} and 
SuperKamiokande \cite{skam} detectors
 reported a deficit in the $\nu_\mu$ flux with
respect to the MC predictions and a distortion of the angular
distributions, while the $\nu_e$ flux and angular distributions agree 
with MC. These features
may be explained in terms of $\nu_\mu \longleftrightarrow
\nu_\tau$ oscillations.

Atmospheric neutrinos are well suited for 
the study
of neutrino oscillations, since they have energies from a fraction of a GeV up 
to more than 100 GeV and they may travel distances $L$ from few tens of km 
up to 13000 km; thus the $L/E_\nu$ values range from $\sim 1$ km/GeV to more
than $10^4$ km/GeV. 

In the following,  we shall briefly recall neutrino oscillation 
formulae  and the main features of the MACRO detector, then we shall discuss
the main results of MACRO, including the ratio of vertical to horizontal 
muons which favour $\nu_\mu \longleftrightarrow
\nu_\tau$ oscillations and the estimates of neutrino energies by multiple
coulomb scattering of muons, which allows the study of the $L/E_\nu$ 
distribution. We shall conclude with a short discussion and by recalling 
briefly other neutrino studies by MACRO.  

\section{Neutrino oscillations}
If neutrinos have non-zero masses, one has to consider 
the neutrino {\it weak flavour eigenstates}
$\nu_e,~\nu_\mu,~\nu_\tau$ and the {\it mass
eigenstates} $\nu_1,~\nu_2,~\nu_3$. 
The weak flavour eigenstates $\nu_l$ are linear combinations of the mass 
eigenstates $\nu_m$ through the elements of the mixing matrix $U_{lm}$:

\begin{equation}
\nu_l = \sum_{m=1}^3 U_{lm}\ \nu_m
\end{equation}
If the mixing angles are small, one would have $\nu_e \sim \nu_1$, 
 $\nu_\mu \sim \nu_2$, $\nu_\tau \sim \nu_3$.
 If the mixing angles are large, the flavour eigenstates are well separated
 from those of mass.

In the simple case of only two flavour eigenstate neutrinos  
$(\nu_\mu,~\nu_\tau)$ which
oscillate with two neutrino mass eigenstates $(\nu_2,~\nu_3)$ one has

\begin{equation}
\left\{ \begin{array}{ll}
      \nu_\mu =~\nu_2 \cos\ \theta_{23} + \nu_3 \sin\ \theta_{23} \\
      \nu_\tau=-\nu_2\sin\ \theta_{23} + \nu_3\cos\ \theta_{23} 
\end{array} 
\right. 
\end{equation} 
\noindent where $\theta_{23}$ is the mixing angle.
 In this case one may easily compute the following expression for the 
survival probability of a $\nu_\mu$ beam (disappearance experiment):
\begin{equation}
P(\nu_\mu \rightarrow \nu_\mu) = 1- \sin^2 2\theta_{23}  
~\sin^2 \left( { {E_2-E_1}\over {2}} t \right) =
1- \sin^2 2\theta_{23}~\sin^2 \left( { 
{1.27 \Delta m^2 \cdot L}\over {E_\nu}} \right)
\end{equation}
where $\Delta m^2=m^2_3-m^2_2$, and $L$ is the distance travelled by the 
neutrino from production to the detection point. The probability 
for the initial $\nu_\mu$ to oscillate into a $\nu_\tau$ is (appearance 
experiment):
\begin{equation}
P(\nu_\mu \rightarrow \nu_\tau) = 1 - P(\nu_\mu \rightarrow \nu_\mu) =
 \sin^2 2\theta_{23}~\sin^2 \left( { 
{1.27 \Delta m^2 \cdot L}\over {E_\nu}} \right)~
\end{equation}

In a disappearance experiment one measures only $P(\nu_\mu \to \nu_\mu)$
as a function of $E_\nu$ and $L$, or of $L/E_\nu$. Only disappearance
experiments have been performed until now. For atmospheric neutrinos 
the parameters
$\theta_{23}$ and $\Delta m^2$ may be determined 
from the variation of
$ P(\nu_\mu \rightarrow \nu_\mu) $ as a function of 
the zenith angle $\Theta$, or from the variation in $L/E_\nu$.
 In future appearance experiments, one would try to observe a neutrino, for
example $\nu_\tau$, not produced in $\pi,~K$ decays.

\section{The MACRO experiment}
\begin{figure}
 \begin{center}
  \mbox{ \epsfysize=7.4cm
	 \epsffile{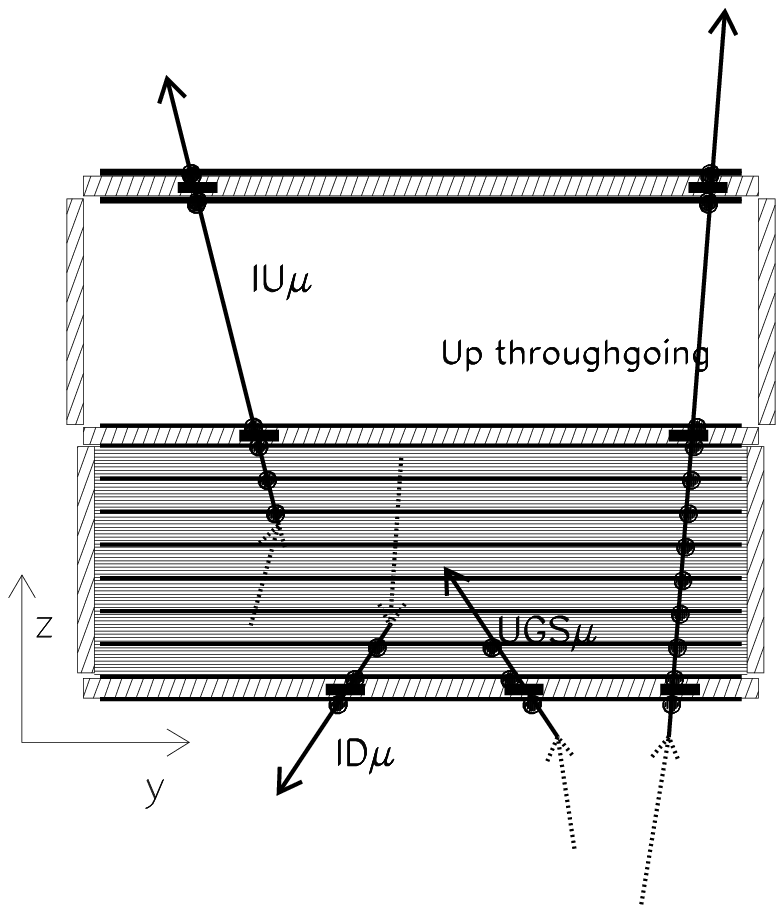}
	\hspace{1cm}
	\epsfysize=5.9cm \epsfxsize=7.5cm
         \epsffile{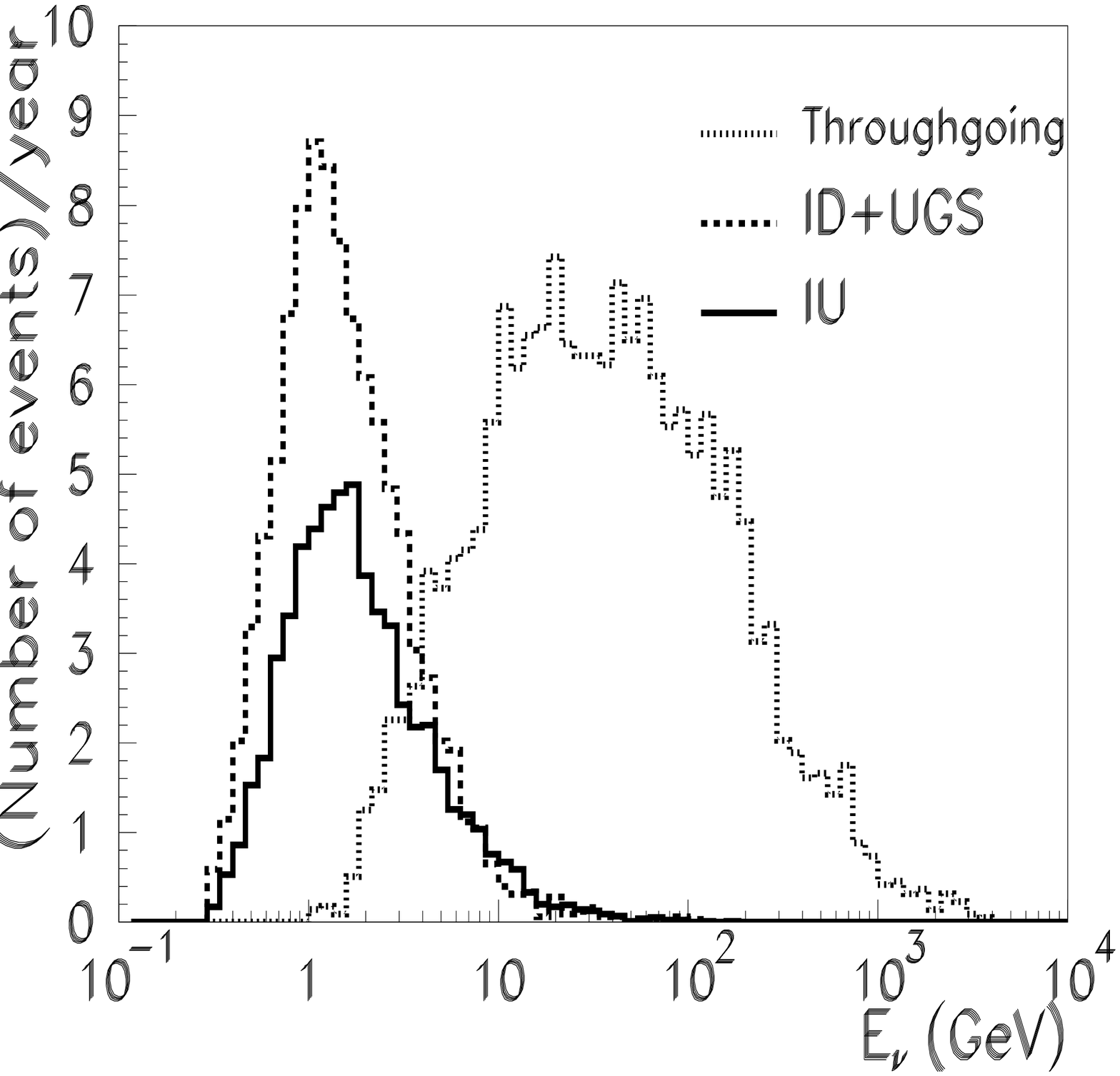} }
 \end{center}
\vspace{-6mm}
 \begin{center}
{\normalsize (a) \hspace{8cm} (b)}
 \end{center}
\vspace{-3mm}
\caption {(a) Vertical 
section of the MACRO detector. Event topologies induced
by $\nu_\mu$ interactions in or around the detector: IU = semicontained 
Internal Upgoing $\mu$; 
 ID = Internal Downgoing $\mu$; UGS = Upgoing Stopping $\mu$; Upthroughgoing =
upward throughgoing $\mu$. The black circles indicate the streamer tube 
hits and the black boxes the scintillator hits. (b) MC simulated distributions 
of the parent neutrino energies, assuming no oscillations, giving rise to 
the different topologies
of muon events detectable by MACRO. The distributions are normalized to one
year of data taking.}
\label{fig:macro}
\end{figure}

The MACRO detector was located in Hall B of the Gran Sasso
Laboratory, under a minimum rock overburden of 3150 hg/cm$^2$ which reduces
the atmospheric muon flux by a factor $\sim 5\cdot 10^5$.
It was a large rectangular box, 76.6m$\times$12m$\times$9.3m, divided
longitudinally in six supermodules and vertically in a lower
part (4.8 m high) and an upper part (4.5 m high) \cite{technical} 
(Fig:~\ref{fig:macro}a).
 It had three types of detectors which gave redundance of informations: liquid
 scintillation counters, limited 
streamer tubes and nuclear track detectors. This last detector was used only
for new particle searches. 

For muon and for neutrino physics and astrophysics studies, 
 the streamer tubes were used for muon tracking 
and the liquid scintillation counters for fast timing. The lower part of
the detector was filled
with trays of crushed rock absorbers alternating with streamer tube
planes; the upper part was open and contained the
electronics. 

There were 10 horizontal planes of streamer tubes in the bottom half of the 
detector, and 4 planes at the
top, all with wire and 27$^\circ$ stereo strip readouts. Six vertical planes
of streamer tubes and one layer of scintillators covered each side of 
the detector.
 The scintillator system consisted of three layers 
of horizontal counters, and of the mentioned  vertical layer
along the sides of the detector. One thus had a closed box structure with
openings only in the upper part of the front and back ends. 

The time and space resolutions for
muons in a scintillation counter were about 500 ps and 11 
cm, respectively, while in the streamer tube system they were about 100 ns
 and 1 cm, respectively. The combination of the informations from the 
streamer tubes and from the scintillators allowed tracking with a precision 
of 1 cm over path lengths of several meters, and timing with a precision of 500 
ps. The detector 
provided a total acceptance $S \Omega \simeq 10000$ m$^2$ sr for an isotropic
flux of particles. 

 Fig:~\ref{fig:macro}a is a vertical section of the detector; it shows 
a general view of the detector and gives also a sketch of the different 
topologies of detected neutrino-induced muon
events used to study neutrino oscillations: Upthroughgoing muons, Internal
Upgoing muons (IU), Upgoing Stopping muons (UGS) and Internal 
Downgoing muons (ID). 
 Fig:~\ref{fig:macro}b shows the parent neutrino
energy distributions obtained from  Monte Carlo calculations for the event 
topologies detectable in MACRO.

\begin{figure}
\vspace{-0.5cm}
 \begin{center}
  \mbox{\epsfysize=6cm
	\epsfxsize=9cm
         \epsffile{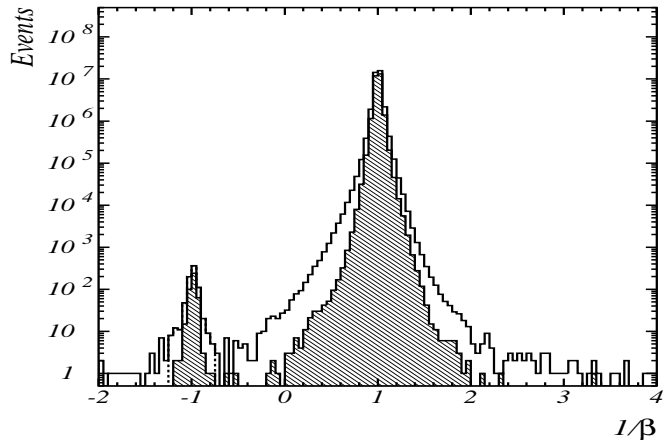}}
 \end{center}
\vspace{-4mm}
\caption{$1/\beta$ distributions for throughgoing (downgoing and upgoing)
 muons for the runs with the complete detector. The shaded 
areas concern muons crossing 3 scintillation counters, while the open
histogram concerns muons crossing 2 or 3 counters. Two vertical dotted lines
delimit the range $-1.25 \leq 1/\beta \leq -0.75$ of upthroughgoing 
muons. There are $\sim 3.5\cdot 10^7$ downgoing muons with $1/\beta \sim 1$ 
and 782 (before subtraction of background) upgoing muons with 
$1/\beta \sim -1$.}
\label{fig:1_su_beta}
\end{figure}

The {\it Upthroughgoing muons} (with $E_\mu > 1$ GeV) come from  
interactions in the rock below the detector of muon neutrinos with an average
energy $\langle E_\nu \rangle \sim 50$ GeV. The tracking is performed 
with streamer tubes 
hits; the time information, provided by scintillation counters,
 allows the determination of the direction (versus) by the time-of-flight 
(T.o.F.) method. If the atmospheric neutrino anomalies are due to neutrino
oscillations, one expects a muon flux reduction depending on the zenith angle.

The {\it semicontained upgoing muons} (IU) come from
$\nu_\mu$ interactions inside the lower apparatus. Since two
scintillation counters are intercepted, the T.o.F. method
is applied to identify the upward going muons. The average 
parent neutrino energy for these events is $\sim 4$ GeV. 
If the atmospheric neutrino anomalies are the results of
$\nu_\mu$ oscillations with maximal mixing
and $10^{-3} < \Delta m^2 < 10^{-2}$ eV$^2$, one expects a 
reduction of about a factor of two in the flux of IU events, 
without any distortion in the shape of the angular distribution.

The {\it up stopping muons} (UGS) are due to $\nu_\mu$ interactions in the rock
below the detector yielding upgoing muon tracks
stopping in the detector; the {\it semicontained downgoing muons} (ID) 
are due to 
$\nu_\mu$ induced downgoing tracks with vertex in the lower MACRO.
The events are found by means of topological criteria; the lack
of time information prevents to distinguish between the two subsamples.
An almost equal number of UGS and ID events is expected.
 In case of oscillations,
 a  similar reduction in the flux of the up stopping events 
and of the semicontained upgoing muons is expected; no reduction is 
instead expected for the semicontained downgoing events (which come from
neutrinos which travelled $\sim 20$ km).

\section{High energy events} 
The data were collected during the running period from  March 1989 
to April 1994 with the detector under construction and during the runs with
the complete 
detector from 1994 until December 2000 (livetime 5.52 yrs). Since the 
total livetime normalized to the full configuration is 6.16 yrs, the
statistics is largely dominated by the full detector run. The analysis of
a data sample of more than 40 million atmospheric downgoing muons
 achieved a rejection factor of $\sim 10^{-7}$ which includes background 
caused by showering events and radioactivity in coincidence with muons.

We studied many sources of possible background: (i) low energy upgoing 
particles (mainly pions) induced by undetected downgoing muons in the rock
surrounding the detector \cite{macro24}. This source of background 
is reduced to 1\% requiring that
more than 200 gr/cm$^2$ of material be crossed in the 
lower part of MACRO; (ii) we 
excluded a region in the azimuthal angle from $-30^\circ$ to $120^\circ$ 
for nearly
horizontal upgoing muons due to insufficient rock overburden; (iii) for muons
 crossing 3 scintillation counters a linear fit of the times as a function of 
the path
length is performed and a cut is applied on the $\chi^2$; (iv) further minor 
cuts are applied to events crossing 2 counters. 

\begin{figure}
 \begin{center}
  \mbox{\hspace{-1cm}
	\epsfysize=7cm
         \epsffile{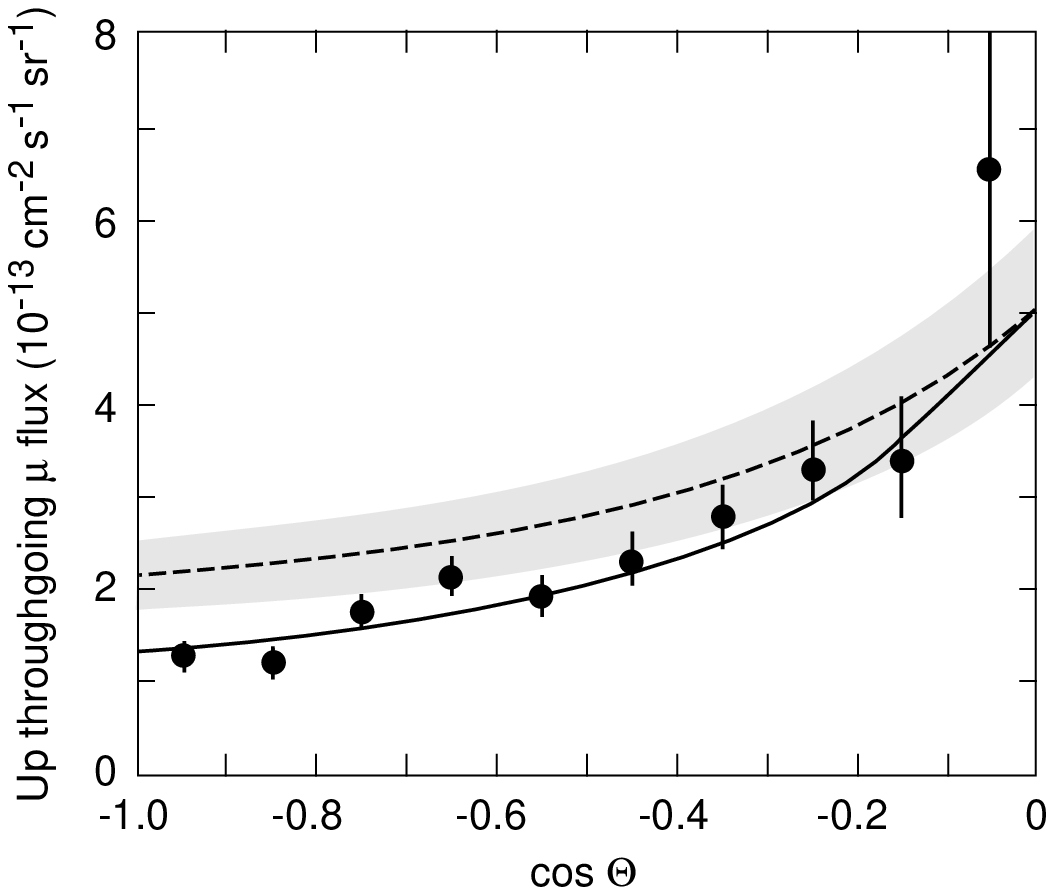}
	\epsfysize=6.8cm \epsfxsize=7.7cm
	\epsffile{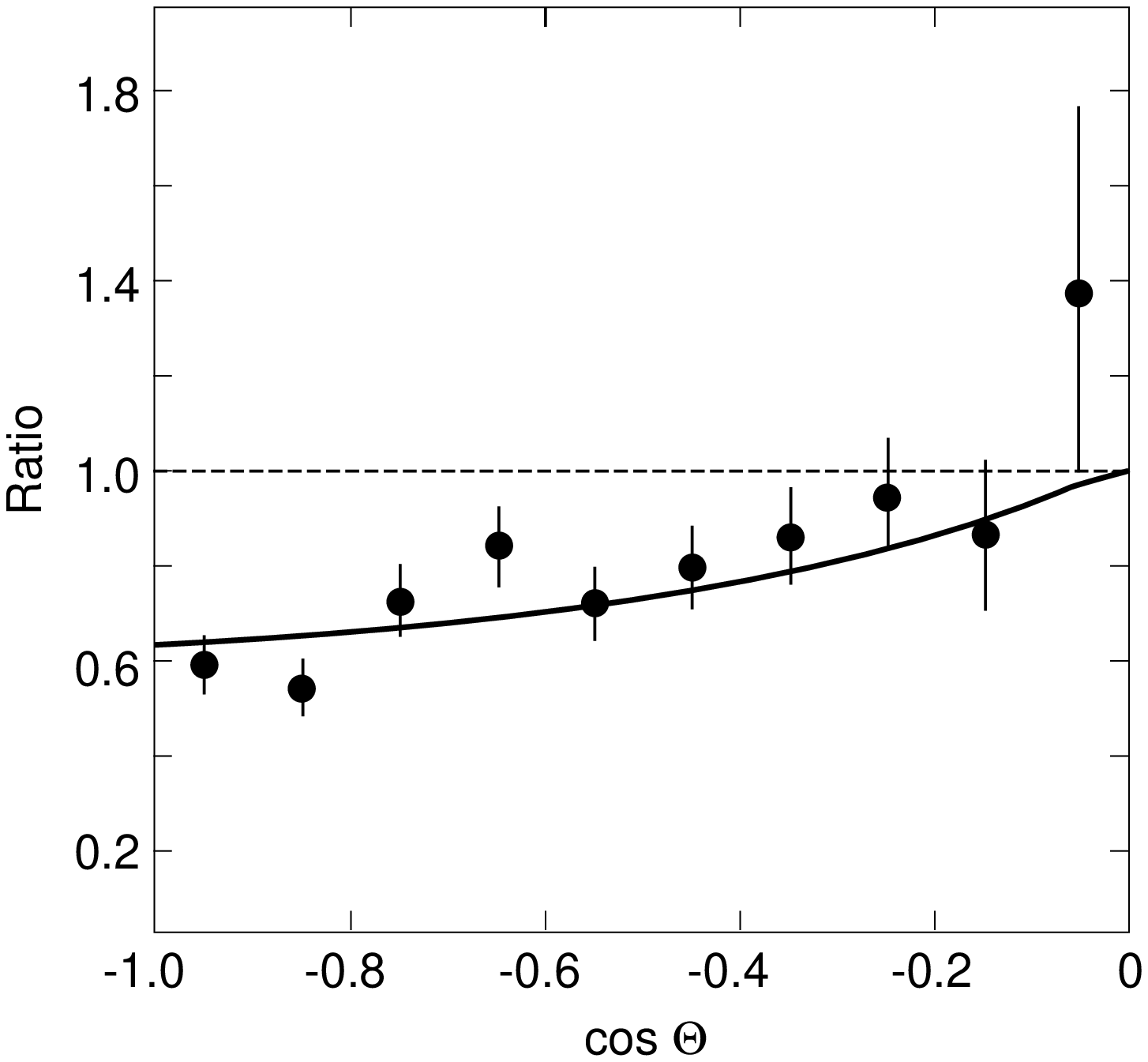}}
 \end{center}
\begin{center}
\vspace{-3mm}
{\normalsize (a) \hspace {7cm} (b)}
\end{center}
\vspace{-3mm}
\caption{(a) Zenith distribution of the upthroughgoing muons (809 
events, background subtracted). The 
data (black points) have statistical and systematic errors
added in quadrature. The shaded region indicates the theoretical scale error 
band of $\pm 17\%$ (see text).
 The solid line is the fit to an oscillated flux with
 maximal mixing and $\Delta m^2 = 2.5\cdot 10^{-3}$ eV$^2$. 
(b) Zenith distribution of the ratio of measured upgoing muons over the
expected ones with no oscillations (black points with statistical and systematic
 errors in quadrature). The solid line is the same ratio computed for 
oscillations; the dotted line corresponds to no oscillations.}
\label{fig:cos-uptr}
\end{figure}

Muons traveling downwards have $1/\beta \sim +1$, upwardgoing muons have 
$1/\beta \sim -1$. The $1/\beta$ distribution for the sample collected with
the full detector is shown in Fig: \ref{fig:1_su_beta}. We selected upwardgoing 
muons requiring $-1.25 \leq 1/\beta \leq -0.75$; we found 863 events, of 
which 809 events remain after subtracting the background.

In the upgoing muon simulation 
the neutrino flux computed by the Bartol group \cite{Agrawal96} was used.
The cross sections for the neutrino interactions was calculated
using the deep inelastic parton distributions  \cite{Gluck95}.
 The muon propagation to the detector was computed using the
energy loss calculation in standard rock \cite{Lohmann85}.
 
The systematic uncertainty arising from the 
neutrino flux, cross section and muon propagation on the
expected flux of upthroughgoing muons was estimated to be $\sim$17\%. 
This systematic uncertainty is
mainly a scale error that doesn't affect the shape of the angular distribution.
 The same cuts applied to the data were used for the simulated events: 
 they selected 1122 MC events, assuming no oscillations.

 Fig: \ref{fig:cos-uptr}a shows the zenith angle $(\Theta$) distribution 
of the measured
flux of upthroughgoing muons of energy $E_\mu > 1$ GeV (black points); the MC 
expectation for no oscillations is indicated by the dashed line with the
shaded scale error band. A 
deficit of events in the region around the vertical can be noticed. 
 The  ratio  of the observed
number of events to the expectation without oscillations 
in $-1 < \cos \Theta < 0$ ($\Theta$ is the zenith angle) is
$0.721 \pm 0.026_{stat} \pm 0.043_{sys} \pm 0.123_{th}$.

We tested the
shape of the angular distribution with the hypothesis of no 
oscillations normalizing
the prediction to the data. The $\chi^2/D.o.F.$ corresponds to a probability
of 0.2\%. In the hypothesis of only $\nu_\mu \longleftrightarrow \nu_\tau$
oscillations, the minimum $\chi^2/D.o.F.$ in the physical region is 
9.7/9 ($P=37$\%)
for maximal mixing and $\Delta m^2 = 2.5\cdot 10^{-3}$ eV$^2$. An
independent test was made on the number of the events. 

 Combining the
probabilities from the two independent tests on the zenith angle shape 
of the flux and on 
the total number of events, the maximal probability is 66\% for maximum 
mixing and 
 $\Delta m^2 \simeq 2.4\cdot 10^{-3}$ eV$^2$. The result of the fit 
is the solid line in Fig:~\ref{fig:cos-uptr}a.

\begin{figure}
 \begin{center}
  \mbox{\epsfysize=6.1cm 
	\epsffile{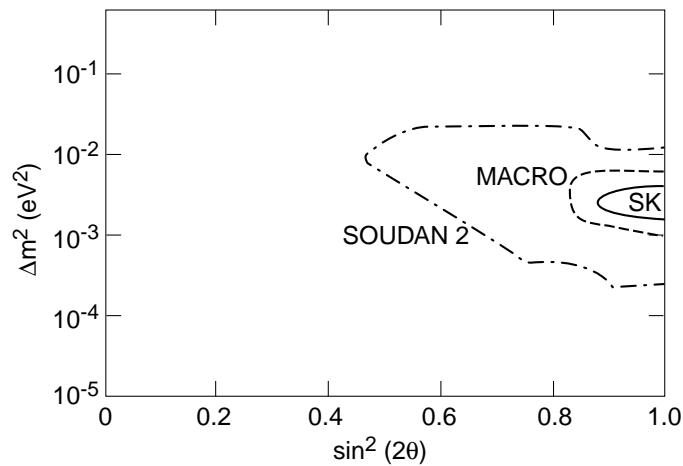}}
 \end{center}
\vspace{-3mm}
\caption{90\% C.L. allowed region contours
for $\nu_\mu \longleftrightarrow \nu_\tau$ oscillations obtained by the 
SuperKamiokande, MACRO and Soudan 2 experiments.}
\label{fig:region}
\end{figure}

For each bin of Fig: \ref{fig:cos-uptr}a we computed the ratio 
$R=$ (measured muon flux)/(expected muon flux in the non oscillation 
hypothesis). The result is shown in  
Fig: \ref{fig:cos-uptr}b (black points). The solid line is the 
ratio $R$ for maximal mixing and 
$\Delta m^2 = 2.5\cdot 10^{-3}$ eV$^2$; the dashed line is the
same ratio expected for no oscillations.  

The MACRO upthroughgoing muons are very sensitive to oscillations because of
the energy range covered. The MACRO 90\% C.L. allowed region in the 
$ \sin^2 2\theta - \Delta m^2$ plane, computed using the Feldman-Cousins 
method \cite{feldman-cousins}, is smaller than the SuperKamiokande (SK) 
one for upthroughgoing muons due to the different energy thresholds
($\sim 1$ GeV for MACRO and $\sim 6$ GeV for SK) and for the
median energies ($\sim 50$ GeV for MACRO, $70\div 80$ GeV for SK).

The MACRO 90\% C.L. allowed region for $\nu_\mu \longleftrightarrow \nu_\tau$
oscillations is compared in Fig: \ref{fig:region} with those obtained by 
the SK and Soudan 2 experiments.

\section{Matter effects. \boldmath $\nu_\mu \longleftrightarrow 
\nu_\tau$ against
$\nu_\mu \longleftrightarrow \nu_{sterile}$} 
Matter effects due to 
the difference between the weak interaction effective potential for 
muon neutrinos with respect to sterile neutrinos (which have null
potential) would produce a different total number and a different zenith 
distribution of upgoing muons. 

\begin{figure}
 \begin{center}
  \mbox{ \epsfysize=6.6cm  
	\epsffile{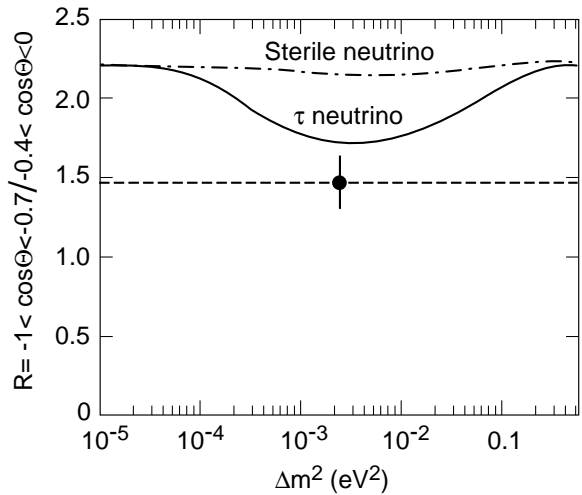}}
 \end{center}
\vspace{-3mm}
\caption {Ratio of events with $-1 < \cos \Theta < -0.7$ to events with
 $-0.4 < \cos \Theta < 0$ as a function of $\Delta m^2$ for maximal mixing.
The black point with error bar is the measured value, the solid line
is the prediction for  $\nu_\mu \longleftrightarrow \nu_\tau$  
oscillations, the dotted line
is the prediction for $\nu_\mu \longleftrightarrow \nu_{sterile}$ 
oscillations.}
\label{fig:nusterile}
\end{figure}

In Fig: \ref{fig:nusterile} the measured ratio 
between the events with $-1 < \cos \Theta < -0.7$ and the events with 
$-0.4 < \cos \Theta < 0$ is shown by the black point. In this ratio 
most of the theoretical uncertainties
on neutrino flux and cross sections cancel. The remaining theoretical 
error, estimated at $\leq 5$\%, has a 3\% contribution 
due to uncertainties on the kaon/pion fraction produced in
atmospheric showers, a 2\% contribution  
from uncertainties in the cross sections
 of almost vertical and almost horizontal
events and a 2.5\% contribution from the seasonal variations of the 
ratio. The systematic experimental error on the 
ratio, due to analysis
cuts and detector efficiencies, is 4.6\%. Combining the experimental
and theoretical errors in quadrature, a global extimate of 7\% is obtained.
MACRO measured 305 events with $-1 < \cos \Theta < -0.7$ and 206 with
$-0.4 < \cos \Theta <0$; the ratio is $R=1.48 \pm 0.13_{stat} \pm 0.10_{sys}$.
For $\Delta m^2=2.5\cdot 10^{-3}$ eV$^2$ and maximal mixing, 
 the minimum expected value of the ratio for 
$\nu_\mu \longleftrightarrow \nu_\tau$ oscillations
is $R_\tau=1.72$; for $\nu_\mu \longleftrightarrow \nu_{sterile}$ oscillations
one expects
$R_{sterile}=2.16$. The maximum probabilities $P_{best}$
to find a value of $R_\tau$ and of $R_{sterile}$ smaller than $R_{expected}$ 
are 9.4\% and 0.06\% respectively. Hence the ratio of the maximum 
probabilities is $P_{best_\tau}/P_{best_{sterile}}=157$, so that 
$\nu_\mu \longleftrightarrow \nu_{sterile}$ oscillations (with any mixing) are 
excluded at 99\% C.L.
compared to the $\nu_\mu \longleftrightarrow \nu_\tau$ channel with 
maximal mixing and $\Delta m^2=2.5\cdot 10^{-3}$ eV$^2$.

\section{Low energy events} 
The data concern only the running period with the detector
in the full configuration 
from April 1994 to December 2000. During this period more than 40 million
downgoing muons were collected. Because of the difference between 
the topologies of the low energy events, two separate analyses were performed.

The IU sample corresponds to an effective livetime of
5.8 yrs. The basic request is the presence of a streamer tube 
track reconstructed in space matching at least two hits in two
different scintillators 
in the upper part of the apparatus. The measured muon 
velocity $\beta c$ was
evaluated with the sign convention that upgoing (downgoing) muons have 
$1/\beta \sim -1~(\sim +1)$. After background subtraction, we had
154 upgoing partially contained events.

The identification of ID+UGS events was based on topological criteria. The
candidates had a track starting (ending) in the lower apparatus and crossing
the bottom detector face. The track has also to be located or oriented in such
a way that it could not have entered (exited) undetected through 
insensitive regions of the apparatus. For this analysis the effective livetime 
was 5.6 yrs. After background subtraction, we had 262 ID+UGS events.

\begin{figure}
 \vspace{-0.3cm}
 \begin{center}
  \mbox{ \epsfysize=6.3cm
        \epsffile{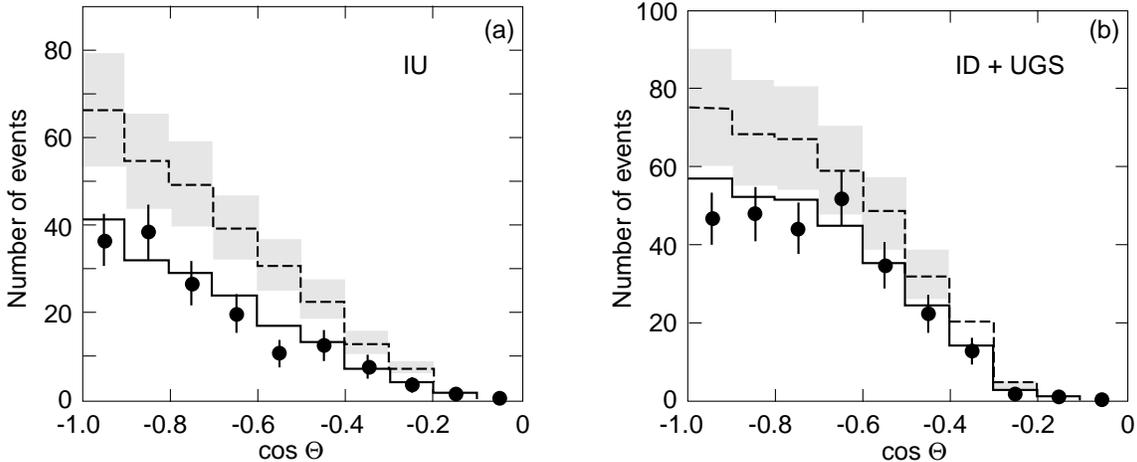}}
 \end{center}
 \vspace{-3mm}
\caption {Measured zenith distributions (a) for the IU events 
  and (b) for the
ID+UGD events. The black points are the data, the shaded regions
correspond to MC predictions assuming no oscillations. The full line is 
the expectation for $\nu_\mu \longleftrightarrow \nu_\tau$ oscillations with
maximal mixing and $\Delta m^2 =2.5\times 10^{-3}$ eV$^2$.}
\label{fig:low_cosze}
\end{figure}

The MC simulations
for the low energy data use the Bartol neutrino
flux \cite{Agrawal96} and the low energy neutrino cross sections 
\cite{lipari94}. We estimated a total theoretical scale uncertainty 
on the predicted number of muons
 of the order of 25\% (it is very probably overestimated). 

{\normalsize
\begin{table}
\begin{center}
\begin{tabular}{cccc}\hline
 & Events & MC$_{\mbox {no osc}}$ & $R=$ (Data/MC$_{\mbox{no~osc}})$ \\
Upthroughgoing & 809 & $1122\pm 191$ & $0.721 \pm 0.026_{stat} 
                   \pm 0.043_{sys} \pm 0.123_{th}$\\
IU      & 154 & $285\pm 28_{sys} \pm 71_{th}$ & $0.54 
                 \pm 0.04_{stat} \pm 0.05_{sys} \pm 0.13_{th}$ \\
ID+UGS  & 262 & $375 \pm 37_{sys} \pm  94_{th}$ & $0.70 
                 \pm 0.04_{stat} \pm 0.07_{sys} \pm 0.17_{th}$ \\
\hline
\end{tabular}
\caption{Summary of the MACRO muon events in $-1 < \cos \Theta < 0$
after background subtraction. For each topology the number of
measured events, the MC prediction for no oscillations and the ratio
$R=$ (Data/MC$_{\mbox{no~osc}})$ are given.}
\end{center}
\label{tab:macro}
\end{table}
}

With the full MC 
simulation, the prediction for IU events was 
$285 \pm 28_{syst} \pm 71_{th}$, while the observed number of events was
$154 \pm 12_{stat}$. The ratio was 
$R_{IU}=$ (Data/MC)$_{IU} = 0.54 \pm 0.04_{stat} \pm 0.05_{syst} 
\pm 0.13_{th}$. For this sample, the average value of $\log_{10} L/E_\nu$
is 3.2, in good agreement with 
the hypothesis of neutrino oscillations.
 
 The prediction for ID+UGS events was $375 \pm 37_{syst} \pm 94_{th}$, while
the observed number of events was $262 \pm 16_{stat}$. The ratio was 
$R_{ID+UGS}=$  (Data/MC)$_{ID+UGS}=0.70 \pm 0.04_{stat} \pm 0.07_{syst} 
\pm 0.17_{th}$.

 The numbers of low energy events are compared with the
predictions in  Table: \ref{tab:macro}; the angular distributions are
shown in Fig: \ref{fig:low_cosze}. The 
 data show a uniform deficit of the measured number of events
over the whole angular distribution with respect to the predictions without
 oscillations.
The two data sets are consistent with neutrino oscillations with 
the parameters found in the analysis of the
high energy sample. Upgoing neutrinos
which induce IU and UGS events, travelling thousands of kilometers, are reduced
by 50\%, while no reduction is expected for downgoing partially contained
muons. As a rough prediction, we expect a rate reduced by 50\% for IU and
by 25\% for ID+UGS.  

 The measured data are in good agreement with the predicted angular 
distributions based on $\nu_\mu \longleftrightarrow \nu_\tau$ 
oscillations with the parameters obtained from the upthroughgoing muon
sample (full histogram in Fig:~\ref{fig:low_cosze}).

\section{\boldmath $\nu_\mu$ energy estimates by Multiple Coulomb 
Scattering of upthroughgoing muons} Since MACRO was not equipped 
with a magnet, 
 the only way to experimentally estimate the muon energy is through 
Multiple Coulomb 
Scattering (MCS) of muons in the absorbers in the lower part of MACRO, 
 see Fig:~\ref{fig:macro}a. The r.m.s. of the
projected displacement of a relativistic muon with momentum $p$ travelling 
for a distance $X$ can be written as:

\begin{equation}
\sigma_{MCS} \simeq \frac{X} {\sqrt{3}} \frac{13.6 \cdot 10^{-3} GeV}
{p\beta c} \sqrt{X/X^0} \cdot (1+0.038 \ln (X/X^0))
\end{equation}
where $p$ (GeV/c) is the muon momentum  and $X/X^0$ is the amount of crossed
material in units of radiation lengths. A muon crossing the whole
apparatus on the vertical has $\sigma_{MCS} \simeq 10 $(cm)/$E$(GeV). The 
muon energy estimate can be performed up to a saturation point, occurring when
$\sigma_{MCS}$ is comparable with the detector space resolution. 

Two analyses were performed. 

The  first 
analysis was made studying the deflection of upthroughgoing muons 
in MACRO with the streamer tubes in digital mode. Using MC methods to estimate
the muon energy from its scattering angle, the data were divided into 3
subsamples with different average energies, in 2 samples in zenith angle 
$\Theta$ and finally in 5 subsamples with different
average values of $L/E_\nu$. This method could reach a spatial resolution of 
$\sim 1$ cm. This analysis yielded an $L/E_\nu$ distribution quite compatible
with neutrino oscillations with the parameters found in Section 4. 

\begin{figure}
 \begin{center}
  \mbox{ \epsfysize=7cm \epsfxsize=9cm
	\epsffile{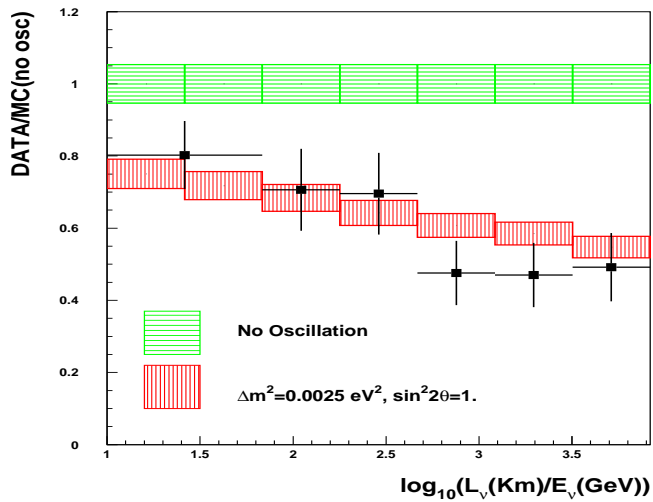}}
 \end{center}
\vspace{-5mm} 
\caption {$R=($Data/MC$_{\mbox {no~osc}})$ as a function of 
$\log_{10} (L/E_\nu)$
obtained using the streamer tubes in ``drift mode".
 The data are the points with error bars, the vertical extent of the shaded 
regions represents the uncertainties on the MC predictions for
$\nu_\mu \longleftrightarrow \nu_\tau$ oscillations and for no oscillations.} 
\label{fig:mcs}
\end{figure}

As the interesting energy region for atmospheric neutrino 
oscillations  spans from $\sim 1 $ GeV to some tens of GeV, it is important
to improve the spatial resolution of the detector to push the saturation
point as high as possible. For this purpose, a second analysis was performed
using the streamer tubes in ``drift mode" \cite{mcs} using the TDC's 
included in the QTP 
system, originally designed for the search for magnetic monopoles. To check
the electronics and the feasibility of the analysis, two ``test beams" were 
performed at the CERN PS-T9 and SPS-X7 beams. For each triggered tube, the 
arrival time of the signal multiplied by the drift velocity, measured at 
the test beam, gives the radius of the drift circle. The
 upthroughgoing muon tracks were reconstructed as the best fit of the
drift circles referring to the triggered streamer tubes. The space resolution
achieved is $\simeq 3$ mm, a factor of 3.5 better than in the 
previous analysis. 
 For each muon, 7 MCS sensitive variables were given in input to a Neural 
Network (NN) which was previously trained with MC events of known input energy 
crossing the detector at different zenith angles. The NN output allows to
separate the whole sample of upthroughgoing muons in 4 subsamples with 
average energies $E_\mu$ of 12, 20, 50 and 102  
GeV, respectively. The comparison of the
zenith angle distributions of the 4 energy subsamples with the predictions
 of no oscillations MC shows a
strong disagreement at low energies (where there is a deficit of vertical 
events), while the agreement is restored at higher 
neutrino energies. The corresponding $\chi^2$ probabilities for the  
no oscillation hypothesis in these energy windows are 
1.8\%, 16.8\%, 26.9\% and 87.7\%, respectively.

The distribution of the ratios $R=($Data/MC$_{\mbox{no~osc}})$ 
obtained by the second method is plotted in Fig: \ref{fig:mcs} as a 
function of $\log_{10} (L/E_\nu)$. The points with error
bars are the data; the vertical extent of the shaded areas represents the  
global uncertainties on the MC predictions $(i)$ for 
$\nu_\mu \longleftrightarrow \nu_\tau$ oscillations with 
maximal mixing and $\Delta m^2 =2.5\times 10^{-3}$ eV$^2$ and $(ii)$ 
for no oscillations.

The 90\% C.L. allowed region from this analysis seems to be 
consistent with the MACRO region plotted in Fig:~\ref{fig:region}.

\section{Conclusions}
We presented and discussed the MACRO results on atmospheric neutrinos. 
 The measured and expected numbers of events are summarized in 
Table:~\ref{tab:macro}. The observed 
zenith angle distributions and the numbers 
of neutrino-induced muons (Fig: \ref{fig:cos-uptr} and 
Fig: \ref{fig:low_cosze}) 
disagree with the predictions of the no oscillation hypothesis.  The 
muon angular distribution and flux are in 
agreement with the hypothesis of two flavour
$\nu_\mu \longleftrightarrow \nu_\tau$ oscillations, with maximal mixing and
$\Delta  m^2 = 2.5\cdot 10^{-3}$ eV$^2$. The hypothesis of 
$\nu_\mu \longleftrightarrow \nu_{sterile}$ oscillations is disfavoured at 
the 99\% C.L. 
The MACRO 90\% C.L. contours in the plane $\sin^2 2 \theta - \Delta m^2$
are compared in Fig. \ref{fig:region} with those obtained by the 
Soudan 2 and SuperKamiokande 
experiments: they overlap closely. The data from the three experiments 
strongly favour $\nu_\mu \longleftrightarrow \nu_\tau$ oscillations, though 
other possible interpretations have been suggested, see for 
example ref. \cite{popa}.

Other MACRO results in the field of neutrino physics and astrophysics 
concern: (i) the
search for cosmic point sources of high energy muon neutrinos: the flux 
upper limits for many sources, like 
Vela P, are at the level of $\Phi_\nu \sim 2\cdot 10^{-6}$ 
cm$^{-2}$ s$^{-1}$ \cite{macro-nuastronomy}; (ii) the search for dark 
matter WIMPs 
annihilating and yielding muon neutrinos from
the earth core and from the Sun: the measured flux upper
limits exclude ranges of parameters of supersymmetric models 
\cite{macro-wimps}; (iii) the 
search for low energy $\overline{\nu_e}$ from stellar gravitational 
collapses: from 1989 till 2000 no type II supernovae 
exploded in our galaxy  \cite{macro-sn}.

\section{Acknowledgements}
We acknowledge the cooperation of many members of the MACRO 
Collaboration, in particular of D. Bakari, G. Battistoni, Y. Becherini, 
 P. Bernardini, T. Montaruli, F. Ronga, E. Scapparone, M. Sioli, M. Spurio, 
 A. Surdo.


\begin{thebibliography}{99}

\bibitem{nulow} 
G. Barr et al., {\it Phys. Rev.} {\bf D39} (1989) 3532; M. Honda 
et al., {\it Phys. Lett.} {\bf B248} (1990) 193; H. Lee and Y.S. 
Koh, {\it Nuovo Cimento} {\bf 105B} (1990) 883.

\bibitem{nuhigh} 
L.V. Volkova, {\it Sov. J. Nucl. Phys.} {\bf 31} (1980) 784; K. Mitsui et al., 
{\it Nuovo Cimento} {\bf 9C} (1986) 995; A.V. Butkevich et al., {\it Sov. 
J. Nucl. Phys.} {\bf 50} (1989) 90.

\bibitem{imb}  IMB Coll., R. Becker-Szendy et al., {\it Phys.
Rev.} {\bf D46} (1992) 3720.

\bibitem{kamioka}
Kamiokande Coll., Y. Fukuda et al., {\it Phys. Lett.} {\bf B335} (1994) 237.

\bibitem{nusex} NUSEX Coll., M. Aglietta et al., 23$^{rd}$  ICRC Proc., 
Calgary, Canada, Vol. {\bf 4} (1993) 446.

\bibitem{frejus} Frejus Coll., K. Daum et al., {\it Z. Phys.} 
{\bf C66} (1995) 417.

\bibitem{baksan} Baksan Coll.,
 S. Mikheyev, 5$^{th}$ TAUP Workshop Proc., Gran Sasso, Italy, 1997.

\bibitem{soudan} Soudan 2 Coll., W.W.M. Allison et al., {\it Phys. 
Lett.} {\bf B391} (1997) 491; {\it Phys. Lett.} {\bf B449} (1999) 137; W. 
Anthony Mann, hep-ex/0007031.

\bibitem{macro}
MACRO Coll., S. Ahlen et al., {\it Phys. Lett.} {\bf B357}(1995) 481. MACRO 
Coll., M. Ambrosio et al., {\it Phys. Lett.} {\bf B434} (1998) 
451; {\it Phys. Lett.} {\bf B478} (2000) 5;  {\it Phys. Lett.} 
{\bf B517} (2001) 59.

\bibitem{skam} SuperKamiokande Coll., Y.Fukuda et 
al., {\it Phys. Rev. Lett.} {\bf 81} (1998) 1562; {\it Phys. Lett.} {\bf B433} 
(1998) 9; {\it Phys. Rev. Lett.} {\bf 85} (2000) 3999; {\it Nucl Phys. B Proc. 
Suppl.} {\bf 91} (2001) 127; T. Toshito, hep-ex/0105023 (2001).

\bibitem{technical} MACRO Coll., M. Ambrosio et al., {\it The MACRO
detector at Gran Sasso}, accepted for publication 
on {\it Nucl. Instr. Meth. A}. 

\bibitem{macro24} MACRO Coll., M. Ambrosio et al., 
{\it Astrop. Phys.} {\bf 9} (1998) 123.

\bibitem{Agrawal96}
V. Agrawal et al., {\it Phys. Rev.} {\bf D53} (1996) 1314.

\bibitem{Gluck95} M. Gluck et al., {\it Z. Phys.} {\bf C67} (1995) 433.

\bibitem{Lohmann85}
W. Lohmann et al., {\it Energy loss of muons in the energy range 1-GeV to 
10000-GeV.}, CERN-EP/85-03 (1985).


\bibitem{feldman-cousins} G.J. Feldman and R.D. 
Cousins, {\it Phys. Rev.} {\bf D57} (1998) 3873.

\bibitem{lipari94}
P. Lipari et al., {\it Phys. Rev. Lett.} {\bf 74} (1995) 4384.

\bibitem{mcs}
 G. Battistoni et al., {\it Neutrino induced upgoing muon energy
estimation by multiple scattering with MACRO}, Cosmic Radiations: From
Astronomy to Particle Physics, NATO Science Series 
II, Vol. 42, p. 141 (2001).

\bibitem{popa} V. Popa and M. Rujoiu, {\it ``Exotic" neutrino 
oscillations}, Cosmic Radiations: From Astronomy to Particle Physics, NATO
Science Series II, Vol. 42, p. 181 (2001).

\bibitem{macro-nuastronomy} MACRO Coll., M. Ambrosio et al., 
{\it ApJ.} {\bf 546} (2001) 1038.

\bibitem{macro-wimps} MACRO Coll., M. Ambrosio et al., {\it Phys. Rev.} 
{\bf D60} (1999) 082002.

\bibitem{macro-sn} MACRO Coll., M. Ambrosio et al., {\it Astrop. Phys.}
{\bf 8} (1998) 123. 

\end{thebibliography}
\end{document}